\documentclass[nofootinbib,preprintnumbers,floatfix,twocolumn,prd,groupedaddress]{revtex4}

\usepackage{amsmath}
\usepackage{amssymb}
\usepackage{graphicx}
\usepackage{slashed}
\usepackage{floatrow}
\usepackage{tikz}
\usepackage{color}
\usepackage{youngtab}
\usepackage{multirow}

\usetikzlibrary{arrows,decorations.pathmorphing,backgrounds,positioning,fit,petri,automata,shadows,calendar,mindmap,
decorations.markings,calc}
\newcommand{\Lag}{\mathcal{L}}

\begin{document}


\title{Renormalization group evolution of dimension-six\\baryon number violating operators}

\author{Rodrigo Alonso}
\author{Hsi-Ming Chang}
\author{Elizabeth E.~Jenkins}
\author{Aneesh V.~Manohar}
\author{Brian Shotwell}

\affiliation{\vspace{1mm}
Department of Physics, University of California at San Diego, La Jolla, CA 92093, USA}


\begin{abstract}

We calculate the one-loop anomalous dimension matrix for the dimension-six baryon number violating operators of the Standard Model effective field theory, including right-handed neutrino fields. We discuss the flavor structure of the renormalization group evolution in the contexts of minimal flavor violation and unification.

\end{abstract}

\maketitle

\section{Introduction}\label{sec:intro}


The baryon asymmetry of the universe hints at baryon number violating (BNV) interactions beyond the Standard Model (SM) of particle physics. Baryon number is an accidental symmetry of the SM violated by quantum effects~\cite{Hooft:1976up}, and there is no fundamental reason why it cannot be violated in extensions of the SM. Indeed, well-motivated theories like grand unified theories~\cite{Georgi:1974sy,Fritzsch:1974nn,Pati:1974yy} violate baryon number at tree level through the exchange of very massive gauge bosons.

There has been no direct experimental observation of baryon number violation to date. The large lower bound for the lifetime of the proton~\cite{Nishino:2009aa,Nishino:2012ipa} requires that the scale of baryon number violation $M_{\slashed{B}}$ be much greater than accessible energy scales, and, in particular, much greater than the SM electroweak scale $M_Z$. The decay of baryons (such as the proton) can then be computed using an Effective Field Theory (EFT) formalism. In the model-independent treatment of EFT, the SM Lagrangian is extended by higher dimensional non-renormalizable operators ($d \geq 5$) suppressed by inverse powers of the new physics scale.

The leading order BNV operators arise at dimension $d=6$. The most general dimension-six Lagrangian can be cast in 63 independent operators~\cite{Buchmuller:1985jz,Grzadkowski:2010es,Weinberg:1979sa,Wilczek:1979hc,Abbott:1980zj}. Out of these 63 operators, 59 operators preserve baryon number, and the complete set of one-loop renormalization group equations for these 59 operators was recently computed in Refs.~\cite{Jenkins:2013zja,Jenkins:2013wua,Alonso:2013hga,Grojean:2013kd}. In the present work, we focus on the four BNV operators~\cite{Weinberg:1979sa,Wilczek:1979hc,Abbott:1980zj}, and we extend the one-loop renormalization group evolution (RGE) analysis to these remaining dimension-six operators.
 
The four BNV operators  can be written\footnote{The connection with the basis of Ref.~\cite{Weinberg:1979sa} is given in Appendix~\ref{sec:appA}.} as~\cite{Abbott:1980zj}
\begin{align}\label{op1-4}
\begin{split}
Q^{duq\ell}_{prst} &= \epsilon_{\alpha\beta\gamma} \epsilon_{ij} (d_p^\alpha C u_r^\beta) (q_s^{i \gamma} C \ell_t^j)\, ,\\
Q^{qque}_{prst} &= \epsilon_{\alpha\beta\gamma} \epsilon_{ij} (q_p^{i \alpha} C q_r^{j \beta}) (u_s^\gamma C e_t)\, ,\\
Q^{qqq\ell}_{prst} &= \epsilon_{\alpha\beta\gamma} \epsilon_{il} \epsilon_{jk} (q_p^{i \alpha} C q_r^{j \beta}) (q_s^{k \gamma} C \ell_t^l)\, ,\\
Q^{duue}_{prst} &= \epsilon_{\alpha\beta\gamma} (d_p^\alpha C u_r^\beta) (u_s^\gamma C e_t) \, ,\\
\end{split}
\end{align}
\noindent where $C$ is the Dirac matrix of charge conjugation, $q$ and $\ell$ are the quark and lepton left-handed doublets, and we use $u$, $d$ and $e$ for up-type, down-type, and charged lepton right-handed fermions. Greek letters denote $SU(3)_c$ color indices and Roman letters from $i$ to $l$  refer to $SU(2)_L$ indices. Roman letters towards the end of the alphabet $p$-$w$ refer to flavor (generation) indices and take on values from $1,\ldots,n_g=3$.

In this work, we also will accommodate neutrino masses for the light neutrinos by including singlet fermions $N$ (right-handed neutrinos) under the SM gauge group. Including singlet $N$ fields, two additional dimension-six BNV operators can be constructed:
\begin{align}\label{op5-6}
\begin{split}
Q^{qqdN}_{prst} &= \epsilon_{\alpha\beta\gamma} \epsilon_{ij} (q_p^{i \alpha} C q_r^{j \beta}) (d_s^\gamma C N_t)\, ,\\
Q^{uddN}_{prst} &= \epsilon_{\alpha\beta\gamma} (u_p^\alpha C d_r^\beta) (d_s^\gamma C N_t) \text{.}\\
\end{split}
\end{align}
The singlet neutrinos $N$, in contrast to the SM fermions, are allowed a Majorana mass $M_N$ by the SM gauge symmetry. $M_N$ can range from a very high scale as in the standard type-I seesaw model~\cite{Minkowski:1977sc,GellMann:1980vs,Mohapatra:1979ia,Schechter:1980gr} to the Dirac neutrino limit for which it vanishes --- see Ref.~\cite{Blennow:2011vn} for a general parametrization in terms of light masses and mixing angles. Even in the case of a very high Majorana mass scale $M_N$, na\"ive estimates of proton decay and light neutrino masses imply that $M_N<M_{\slashed B}$. This hierarchy of scales implies that an EFT with the operators in Eq.~(\ref{op5-6}) holds in the energy regime $M_N<\mu<M_{\slashed B}$. Below the scale $M_N$, one integrates out the $N$ fields, matching onto the EFT containing only the four operators of Eq.~(\ref{op1-4}), and drops the terms of Eq.~(\ref{op5-6}) in the renormalization group equations.

We will use the conventions of Ref.~\cite{Jenkins:2013zja}, generalized to include singlet fermions $N$ \emph{at energies above} $M_N$. Specifically, for $\mu > M_N$, the $\Lag_{d\leq 4}$ SM Lagrangian includes a Majorana mass term $M_N$ for the $N$ fermions as well as Yukawa couplings $Y_N$ for the $N$ and $\ell$ fermions to the electroweak Higgs doublet $H$. For $\mu < M_N$, the $N$ fields are integrated out of the EFT, and $\Lag_{d\leq 4}$ reduces to the conventional SM Lagrangian.

Baryon number is an (anomalous) symmetry that is preserved by the one-loop renormalization group equations, so the dimension-six BNV operators only mix among themselves. The gauge contribution to the anomalous dimensions of Eq.~(\ref{op1-4}) was computed in Ref.~\cite{Abbott:1980zj}, and we agree with those results. In addition, we compute the anomalous dimensions of Eq.~(\ref{op5-6}), and the Yukawa terms. We also classify the operators in terms of representations of the permutation group, which diagonalizes the gauge contributions to the anomalous dimension matrix.

\section{Results}\label{sec:results}
\label{sec:RGE}

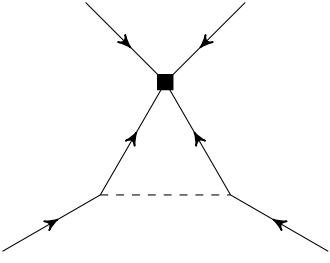
\begin{figure}
\ffigbox{ 
\begin{tikzpicture}
\filldraw (-0.1,0.9) rectangle (0.1,1.1);
\draw[postaction=decorate] [decoration={markings, mark=at position 0.50 with {\arrowreversed[scale=1.5]{stealth'}};}]  (0,1) -- (210:1);
\draw[postaction=decorate] [decoration={markings, mark=at position 0.50 with {\arrowreversed[scale=1.5]{stealth'}};}]  (0,1) -- (330:1);
\draw[dashed] (210:1) -- (330:1);
\draw[postaction=decorate] [decoration={markings, mark=at position 0.50 with {\arrowreversed[scale=1.5]{stealth'}};}]  (90:1) -- +(45:1.5);
\draw[postaction=decorate] [decoration={markings, mark=at position 0.50 with {\arrowreversed[scale=1.5]{stealth'}};}]  (90:1) -- +(135:1.5);
\draw[postaction=decorate] [decoration={markings, mark=at position 0.50 with {\arrowreversed[scale=1.5]{stealth'}};}]  (210:1) -- +(210:1.5);
\draw[postaction=decorate] [decoration={markings, mark=at position 0.50 with {\arrowreversed[scale=1.5]{stealth'}};}]  (330:1) -- +(330:1.5);
\end{tikzpicture}
}{\caption{\label{fig:one-loop} The one-loop Yukawa renormalization graph.}}
\end{figure}

The one-loop anomalous dimension matrix of the BNV operators decomposes into a sum of gauge and Yukawa terms. The gauge anomalous dimension matrix of the operators in Eq.~(\ref{op1-4}) was computed in Ref.~\cite{Abbott:1980zj}. The gauge terms for Eq.~(\ref{op5-6}) have not been computed previously. The Yukawa terms are generated by the diagram in Fig.~\ref{fig:one-loop}, where all the fermion lines are incoming, because of the chiral structure of the BNV operators. The gauge coupling dependence is obtained from an analogous diagram with the scalar replaced by a gauge boson.

The calculation is done using dimensional regularization in $d = 4-2\epsilon$ dimensions in a general $\xi$ gauge. Cancellation of the gauge parameter $\xi$ provides a check on the calculation. The sum of the hypercharges $\mathsf{y}_i$  of the four fermions for each operator is constrained to be equal to zero for the $\xi$-dependence to cancel. Furthermore, the number of colors $N_c=3$ for the operator to be $SU(3)$ gauge invariant. The RGE for the operator coefficients $\mathcal{L}=\sum_i C^i Q^i$ are ($\dot C \equiv 16\pi^2\mu\, {\rm d} C/{\rm d}\mu$):
\begin{widetext}

\begin{align}\label{duqRGE}
\begin{split}
\dot C_{prst}^{duq\ell} =& - C_{prst}^{duq\ell} \left[ 4 g_3^2 + \frac{9}{2} g_2^2 -6(\mathsf{y}_d \mathsf{y}_u + \mathsf{y}_q \mathsf{y}_l) g_1^2 \right] - C_{vrwt}^{duq\ell} (Y_d)_{vs} (Y_d^\dag)_{wp} - C_{pvwt}^{duq\ell} (Y_u)_{vs} (Y_u^\dag)_{wr} \\
& + \left\{ 2C_{prwv}^{duue} + C_{pwrv}^{duue} \right\} (Y_e)_{vt} (Y_u)_{ws}
 - 2 C_{swpv}^{qqdN} (Y_N)_{vt} (Y_u^\dag)_{wr} + \left\{ 2C_{rpwv}^{uddN} + C_{rwpv}^{uddN} \right\} (Y_N)_{vt} (Y_d)_{ws} \\
& + \left\{ 2C_{vwst}^{qqq\ell} + 2C_{wvst}^{qqq\ell} - C_{vswt}^{qqq\ell} - C_{wsvt}^{qqq\ell} + 2C_{svwt}^{qqq\ell} + 2C_{swvt}^{qqq\ell} \right\} (Y_d^\dag)_{vp} (Y_u^\dag)_{wr}+ 2C_{wsrv}^{qque} (Y_d^\dag)_{wp} (Y_e)_{vt}\\
& + C_{vrst}^{duq\ell} (Y_d Y_d^\dag)_{vp} + C_{pvst}^{duq\ell} (Y_u Y_u^\dag)_{vr} + \frac{1}{2} C_{prvt}^{duq\ell} (Y_u^\dag Y_u + Y_d^\dag Y_d)_{vs} + \frac{1}{2} C_{prsv}^{duq\ell} (Y_N^\dag Y_N + Y_e^\dag Y_e)_{vt}
\end{split}
\end{align}
\begin{align}\label{qqueRGE}
\begin{split}
\dot C_{prst}^{qque} =& - C_{prst}^{qque} \left[ 4 g_3^2 + \frac{9}{2} g_2^2 -6(\mathsf{y}_q^2 + \mathsf{y}_u \mathsf{y}_e) g_1^2 \right]- C_{pwvt}^{qque} (Y_u)_{vr} (Y_u^\dag)_{ws} - C_{rwvt}^{qque} (Y_u)_{vp} (Y_u^\dag)_{ws} \\
& + \frac{1}{2} C_{vspw}^{duq\ell} (Y_e^\dag)_{wt} (Y_d)_{vr} + \frac{1}{2} C_{vsrw}^{duq\ell} (Y_e^\dag)_{wt} (Y_d)_{vp} 
 - \frac{1}{2} \left\{ 2 C_{vwst}^{duue} + C_{vswt}^{duue} \right\} \left[ (Y_d)_{vp} (Y_u)_{wr} + (Y_d)_{vr} (Y_u)_{wp} \right] \\
& + \frac{1}{2} \left\{ -2C_{prwv}^{qqq\ell} - 2C_{rpwv}^{qqq\ell} + C_{pwrv}^{qqq\ell} + C_{rwpv}^{qqq\ell} -2C_{wprv}^{qqq\ell} -2C_{wrpv}^{qqq\ell} \right\} (Y_u^\dag)_{ws} (Y_e^\dag)_{vt}\\
& + \frac{1}{2} C_{vrst}^{qque} (Y_u^\dag Y_u + Y_d^\dag Y_d)_{vp} + \frac{1}{2} C_{pvst}^{qque} (Y_u^\dag Y_u + Y_d^\dag Y_d)_{vr} + C_{prvt}^{qque} (Y_u Y_u^\dag)_{vs} + C_{prsv}^{qque} (Y_e Y_e^\dag)_{vt}
\end{split}
\end{align}
\begin{align}\label{qqdRGE}
\begin{split}
\dot C_{prst}^{qqdN} =& - C_{prst}^{qqdN} \left[ 4 g_3^2 + \frac{9}{2} g_2^2 -6\mathsf{y}_q^2 g_1^2 \right] 
 - C_{vrwt}^{qqdN} (Y_d^\dag)_{vs} (Y_d)_{wp} - C_{vpwt}^{qqdN} (Y_d^\dag)_{vs} (Y_d)_{wr}\\
& - \frac{1}{2} C_{swrv}^{duq\ell} (Y_N^\dag)_{vt} (Y_u)_{wp} - \frac{1}{2} C_{swpv}^{duq\ell} (Y_N^\dag)_{vt} (Y_u)_{wr} + \frac{1}{2} \left\{ 2C_{vwst}^{uddN} + C_{vswt}^{uddN} \right\} \left[ (Y_u)_{vp} (Y_d)_{wr} + (Y_u)_{vr} (Y_d)_{wp} \right] \\
& + \frac{1}{2} \left\{ 2C_{prwv}^{qqq\ell} + 2C_{rpwv}^{qqq\ell} - C_{pwrv}^{qqq\ell} - C_{rwpv}^{qqq\ell} +2C_{wprv}^{qqq\ell} +2C_{wrpv}^{qqq\ell} \right\} (Y_d^\dag)_{ws} (Y_N^\dag)_{vt}\\
& + \frac{1}{2} C_{vrst}^{qqdN} (Y_u^\dag Y_u + Y_d^\dag Y_d)_{vp} + \frac{1}{2} C_{pvst}^{qqdN} (Y_u^\dag Y_u + Y_d^\dag Y_d)_{vr} + C_{prvt}^{qqdN} (Y_d Y_d^\dag)_{vs} + C_{prsv}^{qqdN} (Y_N Y_N^\dag)_{vt}
\end{split}
\end{align}
\begin{align}\label{qqqRGE}
\begin{split}
\dot C_{prst}^{qqq\ell} =& - C_{prst}^{qqq\ell} \left[ 4 g_3^2 + 3 g_2^2 -6(\mathsf{y}_q ^2 + \mathsf{y}_q \mathsf{y}_l) g_1^2 \right] - 4 \left\{ C_{rpst}^{qqq\ell} + C_{srpt}^{qqq\ell} + C_{psrt}^{qqq\ell} \right\} g_2^2 \\
& - 4C_{prwv}^{qque} (Y_e)_{vt} (Y_u)_{ws} + 4C_{prwv}^{qqdN} (Y_N)_{vt} (Y_d)_{ws} + 2C_{vwst}^{duq\ell} \left[ (Y_d)_{vp} (Y_u)_{wr} + (Y_d)_{vr} (Y_u)_{wp} \right]\\
& + \frac{1}{2} C_{vrst}^{qqq\ell} (Y_u^\dag Y_u + Y_d^\dag Y_d)_{vp} + \frac{1}{2} C_{pvst}^{qqq\ell} (Y_u^\dag Y_u + Y_d^\dag Y_d)_{vr}
 + \frac{1}{2} C_{prvt}^{qqq\ell} (Y_u^\dag Y_u + Y_d^\dag Y_d)_{vs} + \frac{1}{2} C_{prsv}^{qqq\ell} (Y_N^\dag Y_N + Y_e^\dag Y_e)_{vt}
\end{split}
\end{align}
\begin{align}\label{duuRGE}
\begin{split}
\dot C_{prst}^{duue} =& - C_{prst}^{duue} \left[ 4 g_3^2 -2 \left( 2 \mathsf{y}_d \mathsf{y}_u +2 \mathsf{y}_e \mathsf{y}_u + \mathsf{y}_u^2+ \mathsf{y}_e\mathsf{y}_d \right) g_1^2 \right] +4C_{psrt}^{duue}\left((\mathsf{y}_d +\mathsf{y}_e)\mathsf{y}_u - \mathsf{y}_u^2- \mathsf{y}_e\mathsf{y}_d \right)g_1^2 \\
&+ 4C_{prwv}^{duq\ell} (Y_u^\dag)_{ws} (Y_e^\dag)_{vt} - 8C_{vwst}^{qque} (Y_d^\dag)_{vp} (Y_u^\dag)_{wr}\\
& + C_{vrst}^{duue} (Y_d Y_d^\dag)_{vp} + C_{pvst}^{duue} (Y_u Y_u^\dag)_{vr} + C_{prvt}^{duue} (Y_u Y_u^\dag)_{vs} + C_{prsv}^{duue} (Y_e Y_e^\dag)_{vt}
\end{split}
\end{align}
\begin{align}\label{uddRGE}
\begin{split}
\dot C_{prst}^{uddN} =& - C_{prst}^{uddN} \left[ 4 g_3^2 -2 \left( 2 \mathsf{y}_u \mathsf{y}_d  + \mathsf{y}_d^2 \right) g_1^2 \right] +4C_{psrt}^{uddN}\left(\mathsf{y}_u \mathsf{y}_d - \mathsf{y}_d^2 \right)g_1^2 \\
& + 4C_{rpwv}^{duq\ell} (Y_d^\dag)_{ws} (Y_N^\dag)_{vt} + 8C_{vwst}^{qqdN} (Y_u^\dag)_{vp} (Y_d^\dag)_{wr}\\
& + C_{vrst}^{uddN} (Y_u Y_u^\dag)_{vp} + C_{pvst}^{uddN} (Y_d Y_d^\dag)_{vr} + C_{prvt}^{uddN} (Y_d Y_d^\dag)_{vs} + C_{prsv}^{uddN} (Y_N Y_N^\dag)_{vt}
\end{split}
\end{align}

\end{widetext}

A non-trivial check on these equations is provided by the custodial symmetry limit ($Y_{u(N)}\rightarrow Y_{d(e)},g_1\rightarrow0$).
In order to respect the custodial symmetry, the BNV operator coefficients have to satisfy certain relations given in appendix A, and the RGE flow
should preserve these relations. Remarkably, the construction of custodial invariant operators is compatible with $U(1)_Y$ invariance.

The structure of the anomalous dimensions can be clarified by studying the symmetry properties of the BNV operators. The operators $Q^{qque}$ and $Q^{qqdN}$ are symmetric in the two $q$ indices~\cite{Abbott:1980zj},
\begin{align}\label{qquSymm}
Q_{prst}^{qque} &= Q_{rpst}^{qque}\,, & Q_{prst}^{qqdN} &= Q_{rpst}^{qqdN}\, .
\end{align}
The operator $Q^{qqq\ell}$ satisfies the relation~\cite{Abbott:1980zj},
\begin{align}
Q^{qqq\ell}_{prst} + Q^{qqq\ell}_{rpst} &= Q^{qqq\ell}_{sprt} + Q^{qqq\ell}_{srpt}\,.
\label{reln1}
\end{align}
$Q^{qqq\ell}$ has three $q$ indices, and so transforms like $\Yvcentermath1  \yng(1) \otimes \yng(1) \otimes \yng(1) $, which gives one
completely symmetric, one completely antisymmetric, and two mixed symmetry tensors. Eq.~(\ref{reln1}) implies that one of the mixed symmetry tensors vanishes. The allowed representations of the BNV operators are shown in Table~\ref{tab:1}.

\begin{table*}
\Yvcentermath1
\begin{align*}
\begin{array}{c|c|cccccc}
& \text{dim} & SU(n_g)_q & SU(n_g)_u & SU(n_g)_d & SU(n_g)_l & SU(n_g)_e & SU(n_g)_N  \\[3pt]
\hline
Q^{duq\ell}_{prst} & n_g^4 & \yng(1) & \yng(1) & \yng(1) & \yng(1) & 1 & 1 \\[3pt]
\hline
Q^{qque}_{prst}  & \frac12 n_g^3(n_g + 1) & \yng(2) & \yng(1) & 1 & 1 & \yng(1) & 1 \\[3pt]
\hline
Q^{qqdN}_{prst} & \frac12 n_g^3(n_g + 1)  & \yng(2) & 1 & \yng(1) & 1 & 1 & \yng(1)  \\[3pt]
\hline
\multirow{3}{*}[-14pt]{$Q^{qqq\ell}_{prst}$} & \frac16 n_g^2(n_g + 1)(n_g+2)  & \yng(3)  & 1 & 1 & \yng(1) & 1 & 1\\[7pt]
& \frac13 n_g^2(n_g^2-1)    &  \yng(2,1)  & 1 & 1 & \yng(1) & 1 & 1\\[10pt]
&  \frac16 n_g^2(n_g-1)(n_g-2)  &  \yng(1,1,1) & 1 & 1 & \yng(1) & 1 & 1\\[13pt]
\hline
\multirow{2}{*}[-8pt]{$Q^{duue}_{prst}$} & \frac12 n_g^3(n_g + 1)  & 1 & \yng(2) \ & \yng(1) & 1 & \yng(1) & 1 \\[7pt]
 & \frac12 n_g^3(n_g - 1) & 1 &  \yng(1,1) & \yng(1) & 1 & \yng(1) & 1 \\[7pt]
 \hline
\multirow{2}{*}[-8pt]{$Q^{uddN}_{prst}$} & \frac12 n_g^3(n_g + 1) & 1 & \yng(1) & \yng(2) & 1 & 1 & \yng(1)  \\[7pt]
 &  \frac12 n_g^3(n_g - 1) & 1 & \yng(1) & \yng(1,1) & 1 & 1 & \yng(1)  \\
\end{array}
\end{align*}
\caption{ \label{tab:1} Flavor representations of the BNV operators, and their dimensions.
There are 273 operators in Eq.~(\ref{op1-4}) and  135 in Eq.~(\ref{op5-6}), for a total of 408 $\Delta B=1$ operators with complex
coefficients. One coefficient can be made real by a phase rotation of fields proportional to baryon number. }
\end{table*}

The coefficients $C_{prst}^{duue}$ and $C_{prst}^{uddN}$ can be decomposed into the symmetric and antisymmetric combinations,
\begin{align}\label{qqqSymm2}
C_{prst}^{duue\,(\pm)} &= \frac{1}{2} \left[ C_{prst}^{duue} \pm C_{psrt}^{duue}  \right]\, , \nonumber \\
C_{prst}^{uddN\,(\pm)} &= \frac{1}{2} \left[ C_{prst}^{uddN} \pm C_{psrt}^{uddN}  \right]\, . 
\end{align}
The coefficient $C_{prst}^{qqq\ell}$ can be decomposed into terms with definite symmetry under permutations,
\begin{align}\label{qqqSymm1}
C^{qqq\ell}_{prst} &= S_{prst}^{qqq\ell} + A_{prst}^{qqq\ell} + M_{prst}^{qqq\ell} + N_{prst}^{qqq\ell} \text{,}
\end{align}
\noindent where $S_{prst}^{qqq\ell}$ is totally symmetric in $(p,r,s)$, $A_{prst}^{qqq\ell}$ is totally antisymmetric in $(p,r,s)$, and $M_{prst}^{qqq\ell}$ and $N_{prst}^{qqq\ell}$ have mixed symmetry.

A convenient choice of basis is
\begin{align}\label{qqqSymm2}
S_{prst}^{qqq\ell} &= \frac{1}{6} \left[ C_{prst}^{qqq\ell} + C_{sprt}^{qqq\ell} + C_{rspt}^{qqq\ell} + C_{psrt}^{qqq\ell} + C_{srpt}^{qqq\ell} + C_{rpst}^{qqq\ell} \right]\, , \nonumber \\
A_{prst}^{qqq\ell} &= \frac{1}{6} \left[ C_{prst}^{qqq\ell} + C_{sprt}^{qqq\ell} + C_{rspt}^{qqq\ell} - C_{psrt}^{qqq\ell} - C_{srpt}^{qqq\ell} - C_{rpst}^{qqq\ell} \right]\, , \nonumber \\
M_{prst}^{qqq\ell} &= \frac{1}{3} \left[ C_{prst}^{qqq\ell} - C_{rspt}^{qqq\ell} - C_{rpst}^{qqq\ell} + C_{srpt}^{qqq\ell} \right]\, , \nonumber \\
N_{prst}^{qqq\ell} &= \frac{1}{3} \left[ C_{prst}^{qqq\ell} - C_{sprt}^{qqq\ell} + C_{rpst}^{qqq\ell} - C_{srpt}^{qqq\ell} \right] \text{.}
\end{align}
\noindent The coefficient $M_{prst}^{qqq\ell}$ is obtained by first anti-symmetrizing $C_{prst}^{qqq\ell}$ in $(p,r)$, and then symmetrizing in $(p,s)$. Likewise, $N_{prst}^{qqq\ell}$ is obtained by first anti-symmetrizing in $(p,s)$, and then symmetrizing in $(p,r)$. Eq.~(\ref{reln1}) implies that $N_{prst}^{qqq\ell} $ vanishes.

The gauge contributions to the anomalous dimensions respect the flavor symmetry of the operators. With the decomposition Eq.~(\ref{qqqSymm2}), the gauge contribution to the anomalous dimension matrix diagonalizes,
\begin{align}\label{qqqSymm3}
\dot C_{prst}^{duue\,(\pm)} &= -\left[4g_3^2+\left(2 \pm  \frac{20}{3}\right)g_1^2 \right] C_{prst}^{duue\,(\pm)} + \ldots  \nonumber \\
\dot C_{prst}^{uddN\,(\pm)} &= -\left[4g_3^2+\left(\frac23 \pm  \frac{4}{3}\right)g_1^2 \right] C_{prst}^{uddN\,(\pm)} + \ldots   \nonumber \\
\dot S_{prst}^{qqq\ell} &= -\left[4g_3^2+15g_2^2+\frac13g_1^2 \right] S_{prst}^{qqq\ell} + \ldots  \nonumber \\
\dot A_{prst}^{qqq\ell} &= -\left[4g_3^2-9g_2^2+\frac13g_1^2 \right] A_{prst}^{qqq\ell} + \ldots  \nonumber \\
\dot M_{prst}^{qqq\ell} &= -\left[4g_3^2+3g_2^2+\frac13g_1^2 \right] M_{prst}^{qqq\ell} + \ldots \, .
\end{align}
The ``$\,\cdots$'' refers to the Yukawa contributions, which can mix different permutation representations.

\section{Discussion}\label{sec:disc}

The renormalization group equations presented here have an involved flavor structure; to better understand the generic features, we
turn now to certain simplifying hypotheses and models that produce a simple subclass of BNV operators.
\subsection{Minimal Flavor Violation}\label{subsec:mfv}

The SM has an $SU(3)^5$ flavor symmetry for the $q, u, d, l$, and $e$ fields, broken only by the Higgs Yukawa interactions. The symmetry is preserved if we promote the Yukawa coupling matrices to spurions that transform appropriately under the flavor group. Minimal flavor violation (MFV)~\cite{SekharChivukula198799,DAmbrosio:2002ex} is the hypothesis that any new physics beyond the SM preserves this symmetry, so the Yukawa coupling matrices are the only spurions.

Dimension-six BNV operators do not satisfy na\"ive minimal flavor violation because of triality. The argument proceeds as follows: under every $SU(3)_i$ flavor transformation, each BNV operator transforms as a representation of $SU(3)_i$ with $n_i$ upper indices and $m_i$ lower indices. All BNV operators satisfy $\sum_{i=1}^5 (n_i-m_i) \equiv 1$ (mod 3). No combination of Yukawa matrices (or other invariant tensors) can change this into a singlet, as they all have $(n-m) \equiv 0$ (mod 3).

In extensions of the MFV hypothesis to account for massive neutrinos~\cite{Cirigliano:2005ck,Davidson:2006bd,Alonso:2011jd}, a Majorana mass term introduces a spurion with $(n-m) \equiv 2$ (mod 3). This in turn allows for the implementation of MFV, as pointed out in Ref.~\cite{Nikolidakis:2007fc}. Note also that if the Yukawa spurions are built out of objects with simpler flavor-transformation properties~\cite{Alonso:2011yg}, a variant of minimal flavor violation is possible without Lepton number violation.


Finally, there is the possibility that the fermion fields do not each separately have an $SU(3)$ flavor symmetry, but that some transform simultaneously~\cite{Grinstein:2006cg}. The latter is an attractive option that is realized in Grand Unified Theories (GUTs), and we explore this possibility in the next subsection.

\subsection{Grand Unified Theories}\label{subsec:guts}

The Georgi-Glashow $SU(5)$ theory~\cite{Georgi:1974sy} places $u^c$, $q$, and $e^c$ in a ${\bf 10}$ representation of $SU(5)$, and $d^c$ and $l$ in a ${\bf \overline{5}}$. In the context of the type-I seesaw, $N$ is a ${\bf 1}$. The flavor group in this case cannot be that of putative MFV since the fields in each $SU(5)$ representation must transform simultaneously. The flavor symmetry is instead $SU(3)^3 = SU(3)_{\bf 10} \otimes SU(3)_{\bf \bar 5} \otimes SU(3)_{\bf 1}$, where each $SU(3)$ stands for transformations in flavor space of the corresponding $SU(5)$ representation~\cite{Grinstein:2006cg}. The fermions and spurions then fall into the representations
\begin{equation}\label{su5mfv_field}
\begin{alignedat}{2}
u^c, q, e^c \sim & \ ({\bf 3,1,1})\, , \quad{} \quad{} & Y_u \sim & \ ({\bf \bar{6},1,1})\, , \quad{} \\
d^c, l \sim & \ ({\bf 1,3,1})\, , \quad{} \quad{} & Y_d, Y_e^T \sim & \ ({\bf \bar{3},\bar{3},1})\, , \quad{} \\
N^c \sim & \ ({\bf 1,1,3})\, , \quad{} \quad{} & Y_N \sim & \ ({\bf 1,\bar{3},\bar{3}})\, , \quad{} \\
& \quad{} \quad{} & M_N \sim & \ ({\bf 1,1,6})\, , \quad{} \\
\end{alignedat}
\end{equation}

\noindent where the right-handed neutrino Majorana mass $M_N$ also needs to be promoted to a spurion. Note that the triality argument given previously does not apply to the Yukawa matrices in this scenario. With the $SU(5)$ GUT in mind, we will relabel the Yukawas $Y_u \to Y_{10}$, $(Y_d, Y_e^T) \to Y_5$, and $Y_N \to Y_1$.

The operators transform as
\begin{align}\label{su5mfv_op}
Q^{duq\ell} &\sim  \ ({\bf 3 \otimes \bar{3}, 3 \otimes \bar{3}, 1}) , \nonumber\\
Q^{qqq\ell} & \sim  \ ({\bf 3 \otimes 3 \otimes 3, 3, 1}),\nonumber\\
Q^{uddN} &\sim  \ ({\bf \bar{3}, \bar{3} \otimes \bar{3}, \bar{3}}),  \nonumber\\
Q^{duue} & \sim  \ ({\bf \bar{3} \otimes \bar{3} \otimes \bar{3}, \bar{3}, 1}),\nonumber\\
Q^{qqdN} &\sim  \ ({\bf 3 \otimes 3, \bar{3}, \bar{3}}), \nonumber \\
Q^{qque} &\sim  \ ({\bf 3 \otimes \bar{3} \otimes 3 \otimes \bar{3},1, 1}),
\end{align}
which now can be combined with Yukawa couplings to build up invariant terms in the Lagrangian. Explicitly, the coefficients of the operators in terms of Yukawa matrices up to second order are
\begin{align}\label{su5mfv_op}
C^{duq\ell} &\sim \ 1 \oplus Y_{10}^\dag Y_{10} \oplus Y_5^\dag Y_5\, ,  \nonumber\\
C^{qqq\ell} &\sim  \ Y_{10} \otimes Y_5\, ,  \nonumber \\
C^{uddN} & \sim  \ Y_5^\dag \otimes Y_1^\dag\, ,\nonumber\\
C^{duue} & \sim  \ Y_{10}^\dag \otimes Y_5^\dag\, ,  \nonumber \\
C^{qqdN} &\sim  \ Y_{10} \otimes Y_1^\dag\, , \nonumber \\
C^{qque} &\sim  \ 1 \oplus Y_{10} Y_{10}^\dag \oplus Y_{10} \otimes Y_{10}^\dag\, .
\end{align}

Notice that only $C^{duq\ell}$ and $C^{qque}$ can be constructed out of flavor singlets. These are the only two operators that can be generated by integrating out heavy gauge bosons in the context of $SU(5)$ or, in general, by flavor-blind $SU(5)$ invariant dynamics. In addition, these are the only two coefficients that remain in the limit $Y_5, Y_1 \to 0$ ($Y_d, Y_e, Y_N \to 0$).

To close this section, let us comment on the implications for supersymmetric GUTs in our framework. BNV dimension-five operators are produced by integrating out GUT particles in supersymmetric theories in the absence of selection rules like $R$-parity~\cite{Sakai:1981pk,Dimopoulos:1981dw,Weinberg:1981wj}. Below the supersymmetry breaking scale, these will translate into the operators $Q^{qqq\ell}$, $Q^{duue}$ and $Q^{uddN}$ in terms of the SM EFT Lagrangian, being only suppressed by one power of the BNV scale: $1/(M_{\slashed{B}} M_{\text{SUSY}})$. A feature of this scenario is that, as a result of the supersymmetric origin of the operators, all diagonal entries in flavor vanish~\cite{Dimopoulos:1981dw}, so that proton decay would require a strange particle. The renormalization group equations presented here only apply in the regime $\mu<M_{\text{SUSY}}$ since they depend on the spectrum of the theory, and we have assumed only dynamical SM particles. See Ref.~\cite{Bernon:2014fpa} for a RGE study of BNV effects in the context of supersymmetry.

\subsection{Magnitude of Effects}\label{subsec:effects}

In this subsection, we simplify the RGE to estimate the magnitude of running a BNV operator coefficient from the GUT scale to the electroweak scale. Working in the context of a MFV GUT discussed in Sec.~\ref{subsec:guts}, we set $Y_d = Y_e = Y_N = 0$, assuming top-Yukawa dominance. In that limit, the only two non-vanishing operators are $Q_{prst}^{duq\ell}$ and $Q_{prst}^{qque}$, whose RGE equations decouple. The coefficients of these two operators are given by appropriate combinations of $Y_{10}$ which transforms as the symmetric representation, $\bf\bar6$.

As an example, we focus on $Q^{duq\ell}_{prst}$, whose coefficient takes on a simple form:
\begin{align}\label{Op1Simp}
C_{prst}^{duq\ell} &= C_{rs}^{duq\ell} \delta_{pt}, \ \ \text{where} \ \ C_{rs}^{duq\ell} = f(Y_{10}^\dag Y_{10})_{rs}\,,
\end{align}
\noindent and $f(0)_{rs} \propto \delta_{rs}$. The RGE of this coefficient becomes
\begin{align}\label{Op1SimpRGE}
\begin{split}
\dot C_{rs}^{duq\ell} \to \left[ \frac{1}{2} Y_{10}^\dag Y_{10} - 4g_3^2 - \frac{9}{2}g_2^2 - \frac{11}{6}g_1^2  \right]_{rw} \! \! C_{ws}^{duq\ell} \text{.}
\end{split}
\end{align}

We can now choose the basis $Y_{10} = Y_u = \text{diag} (0,0,y_t)$, where $y_t$ is the top-quark Yukawa coupling and lighter up-type quark masses are neglected. With this simplification, $C_{rs}^{duq\ell}$ is a diagonal matrix. Setting $M_{GUT} \approx 10^{15}\,$GeV, the $C^{duq\ell}$ coefficients at the electroweak and GUT scales are related by
\begin{align}\label{CduqRunning}\nonumber
C^{duq\ell}_{33}(M_Z) \approx & \ (2.26)(0.96) \, C^{duq\ell}_{33}(M_{\text{GUT}}) \,,\\
C^{duq\ell}_{22\,(11)}(M_Z) \approx & \ (2.26) \, C^{duq\ell}_{22\,(11)}(M_{\text{GUT}}) \,.
\end{align}
\noindent The first factor in parentheses comes from the gauge contribution alone,  is dominated by the QCD coupling, and is common to all flavor coefficients. The second factor is the extra correction from including the Yukawa contribution, with only the top entry sizeable. Whereas the gauge contribution to the RGE enhances the $C_{rs}^{duq\ell}$ coefficient at lower energy scales, the Yukawa contribution gives a small suppression.

The Yukawa-induced running will in general be negligible for the lightest generation coefficients and processes like proton or neutron decay are unaffected. The Yukawa running gives a small correction for heavier generations. Note that the relatively small correction from Yukawa running compared to gauge-induced running stems from the different numerical coefficients of the anomalous dimension, since $g_3 \sim y_t$. For example, in Eq.~(\ref{Op1SimpRGE}), the color and $SU(2)_L$ gauge contributions have \emph{each} a pre-factor $\sim\!8$ times that of the Yukawas. These numerical factors cannot be estimated and require the explicit computation presented here.

The Yukawa running studied in this section have the most impact in heavy flavor BNV transitions, which are searched for experimentally~\cite{BABAR:2011ac,Chatrchyan:2013bba}. In this regard, the fact that $W$ boson exchange below the electroweak symmetry-breaking scale produces flavor mixing is relevant. In particular, at two-loop order, proton or neutron decay is sensitive to BNV operators with arbitrary flavor. Even though a two-loop effect, this places a  strong bound on heavy flavor BNV. Discussions of heavy BNV transitions taking into account these effects can be found in Refs.~\cite{Morrissey:2005uza,Hou:2005iu,Dong:2011rh}.

\section{Conclusions}\label{sec:conc}
 
In this letter, we have included the Yukawa contribution to the anomalous dimension matrix of baryon number violating operators and have thus completed the one-loop renormalization group evolution. Together with the computation of Refs.~\cite{Jenkins:2013zja,Jenkins:2013wua,Alonso:2013hga}, this completes the anomalous dimension matrix for the totality of dimension-six operators of the SM. We included right-handed neutrinos and therefore two new BNV operators, and classified all the operators under flavor symmetry. None of the operators satisfies $SU(3)^5$ minimal flavor violation, but it is possible to impose a weaker grand unified theory variant of MFV. The Yukawa coupling corrections only give small corrections to the operator evolution.

\begin{acknowledgements}
This work was supported in part by DOE grant DE-SC0009919.
\end{acknowledgements}

\appendix

\section{Operator Relations and Custodial Symmetry}\label{sec:appA}

Refs.~\cite{Weinberg:1979sa,Grzadkowski:2010es} split the $Q^{qqq\ell}$ operator into two operators
\begin{align}\label{qqq13}
\begin{split}
Q^{qqq\ell\,(1)}_{prst} &= \epsilon_{\alpha\beta\gamma} \epsilon_{ij} \epsilon_{kl} (q_p^{i \alpha} C q_r^{j \beta}) (q_s^{\gamma k} C l_t^l)\, ,\\
Q^{qqq\ell\,(3)}_{prst} &= \epsilon_{\alpha\beta\gamma} (\tau^I \epsilon)_{ij} (\tau^I \epsilon)_{kl} (q_p^{i \alpha} C q_r^{j \beta}) (q_s^{\gamma k} C l_t^l) \, , \\
\end{split}
\end{align}
\noindent where $\tau^I$ is an $SU(2)_L$ generator. These operators can be written in terms of $Q^{qqq\ell}_{prst}$~\cite{Abbott:1980zj}
\begin{align}\label{qqq13fromqqq}
\begin{split}
Q^{qqq\ell\,(1)}_{prst} &= -(Q^{qqq\ell}_{prst} + Q^{qqq\ell}_{rpst})\, ,\\
Q^{qqq\ell\,(3)}_{prst} &= -(Q^{qqq\ell}_{prst} - Q^{qqq\ell}_{rpst})\, ,\\
\end{split}
\end{align}
$Q^{qqq\ell\,(1)}_{prst}$ and $Q^{qqq\ell\,(3)}_{prst}$ are symmetric and antisymmetric in the first two flavor indices, respectively, and transform as symmetric plus mixed, and antisymmetric plus mixed representations under permutation of the three $q$ indices. Since there is only one mixed symmetry tensor in $Q^{qqq\ell}$ by Eq.~(\ref{reln1}), the mixed symmetry tensors in $Q^{qqq\ell\,(1,3)}$ are the same, and the two operators are not independent.

The custodial $SU(2)_L\times SU(2)_R$ symmetry is preserved in the SM for $g_1 \to 0$ and $Y_{u(N)} \to Y_{d(e)}$. It can be implemented in the BNV operators by arranging the right-handed fermions in doublets, $q_R=(u_R,d_R)^T$ and $\ell_R=(N_R, e_R)^T$. By construction, $Q^{qqq\ell}$ is already custodial invariant and the five remaining operators are grouped into the custodial $SU(2)$ invariant combinations
\begin{align}
\begin{split}
\epsilon_{ij} \epsilon_{kl} (q^i_{R\,p} C q^j_{R\,r}) (q_s^{ k} C \ell_t^l)&=-Q^{duq\ell}_{prst}-Q^{duq\ell}_{rpst}\, ,\\
\epsilon_{ij} \epsilon_{kl} (q_p^{i} C q_r^{j}) (q_{R\,s}^{ k} C \ell_{R\,t}^l)&=Q^{qque}_{prst}-Q^{qqdN}_{prst}\, ,\\ 
\epsilon_{ij} \epsilon_{kl} (q_{R\,p}^{i} C q_{R\,r}^{j}) (q_{R\,s}^k C \ell_{R\,t}^l)&=-Q^{uddN}_{prst}-Q^{uddN}_{rpst}\\
&\quad-Q^{duue}_{prst}-Q^{duue}_{rpst}\, ,
\end{split}
\end{align}
\noindent where color indices are implicit.
The component fields of $q_R$ and $\ell_R$ have different hypercharges, but the custodial invariant operators \emph{are} $U(1)_Y$ invariant. The above equations imply extra relations for the operator coefficients
\begin{align}\label{custSU2rel}\nonumber
C_{prst}^{duq\ell} =& C_{rpst}^{duq\ell}, &
C_{prst}^{qque} =& -C_{prst}^{qqdN},  \\
C_{prst}^{duue}=&C_{rpst}^{duue}, &C_{prst}^{duue} =& C_{prst}^{uddN}\,,
\end{align}
in the custodial $SU(2)$ limit.

\bibliography{BNV}

\end{document}